\documentclass[preprint,aps,pre]{revtex4-1} 
\usepackage{amssymb}
\usepackage{graphicx}
\usepackage{color}
\usepackage{float}  
\usepackage{amsmath}

\usepackage{makeidx}  
\usepackage{ifpdf}

\usepackage{url,hyperref}
\usepackage[usenames,dvipsnames]{xcolor}
\hypersetup{colorlinks=true, linkcolor=BrickRed, urlcolor=blue!50!black, citecolor=blue!50!black}
\usepackage[capitalize]{cleveref}
\usepackage{cleveref}
\crefrangelabelformat{equation}{(#3#1#4)$-$(#5#2#6)}
\begin{document}

\title{Macroscopic rotation of active colloids in a colloid-polymer mixture confined inside a spherical cavity}
 
\author{Arabinda Bera$^{1}$, Kurt Binder$^{2}$, Sergei A. Egorov$^{3}$, and Subir K. Das$^{1}$}
\email{das@jncasr.ac.in}

\affiliation{$^1$Theoretical Sciences Unit and School of Advanced Materials, Jawaharlal Nehru Centre for Advanced Scientific Research, Jakkur P.O, Bangalore 560064, India}

\affiliation{$^2$Institut f\"{u}r Physik, Johannes Gutenberg-Universit\"{a}t, D-55099 Mainz, Staudinger Weg 7, Germany}

\affiliation{$^3$Department of Chemistry, University of Virginia, Charlottesville, Virginia 22901, USA}

\date{\today}
\begin{abstract}
From studies via Molecular Dynamics simulations, we report results on structure and dynamics in mixtures of active colloids and passive polymers that are confined inside a spherical container with a repulsive boundary. Such systems mimic the presence of bacteria in the background of bio-polymers. All interactions in the fully passive case are chosen in such a way that in equilibrium coexistence between colloid-rich and polymer-rich phases occurs. For most part of the studies the chosen compositions give rise to Janus structure; one side of the sphere is occupied by the colloids and the rest by the polymers. This partially wet situation mimics nearly a neutral wall in the fully passive scenario. Following the introduction of a velocity-aligning activity to the colloids, the shape of the polymer-rich domain changes to that of an ellipsoid, around the long axis of which the colloid-rich domain attains a macroscopic angular momentum. In the steady state, the orientation of this axis evolves via diffusion, implying that the passive domain is set into motion as well by the active particles.
\end{abstract}

\maketitle
\section{Introduction}
Understanding how phase transitions and collective behavior of particles in active matter differ from their equilibrium counterparts in passive systems has become a grand challenge problem \cite{gompper2020,marchetti2016,bechinger2016,cates2015,marchetti2013}. A key observation, dating back to the Vicsek \cite{vicsek1995} model for the formation of bird swarms, is the ``motility induced phase separation'' (MIPS) \cite{marchetti2016,cates2015,solon2018} between high-density (liquid-like) and low density (vapor-like) phases. Experimentally, MIPS was found in suspensions of motile bacteria \cite{liu2019} or of self-propelled colloids \cite{buttinoni2013}, for instance.

As a ``fruitfly model'' of MIPS, the model of active Brownian particles (ABP) was proposed \cite{fily2012,speck2014,mallory2018,speck2021}. These particles interact via pairwise repulsive forces in the absence of hydrodynamics, each particle having an intrinsic speed of fixed absolute value $v_0$, with the direction decorrelating by rotational diffusion. Unlike Ref. \cite{vicsek1995}, there is no explicit alignment interaction in this model.

Recent studies \cite{turci2021,omar2021} revealed that the phase behavior of this ABP model is apparently very complicated in space dimension $d=3$: liquid-vapor-like phase coexistence typically is metastable against separation between vapor-like and crystalline phases. Thus, we focus here on an alternative model \cite{das2014,trefz2016,trefz2017,binder2021}, viz., an Asakura-Oosawa (AO) type model for colloid-polymer mixtures \cite{asakura1954,vrij1976,binder2014} amended by a Vicsek-like interaction, the colloids being self-propelling. Choosing a size ratio $q_s=0.8$ between the radius of the polymer coils (that are modelled as soft spheres) and the colloids, the phase behavior in the passive case has been precisely characterized: a vapor-liquid-like separation occurs, with  Ising critical behavior \cite{binder2014,vink2004,vink2005,zausch2009} remaining unaffected.

A strange feature of the simulation studies of the above mentioned active AO model \cite{das2014,trefz2016,trefz2017} is the anomalous behavior at the interfaces between the coexisting phases. This is enforced by the periodic boundary conditions (PBC) of the cubic box, standard for obtaining the bulk behavior of quasi-macroscopic systems \cite{allen2017}: the slab-like liquid domain is separated from the vapor by planar interface at which the ``local temperature'' is enhanced by a factor of 2 to 4 compared to the rest of the system \cite{trefz2016}. Here the temperature is defined kinetically in terms of the mean-squared fluctuation of the particle velocities. No such effects occur in equilibrium. There the temperature is homogeneous, and PBC do not cause such unphysical artefacts. Thus, small finite systems with PBC are well suited to elucidate the bulk behavior of real systems in equilibrium, while the use of PBC can cause problems for simulations of active matter. 

A physically more reasonable behavior is expected for a system in a closed container confined by walls. The equilibrium behavior of the AO model in spherical confinement has revealed interesting core-shell or Janus-particle type structures, depending on the particle-wall interactions \cite{winkler2013}. Thus, to investigate the interesting case of active matters enclosed in vesicles or living cells, we study here an active version of the AO model \cite{das2014,trefz2016,trefz2017,binder2021} under spherical confinement, where, as stated before, the colloids are made self-propelling. We observe that the region of phase coexistence between colloid-rich and polymer-rich phases in the confined system becomes wider with increasing ``activity strength'' $f_A$: this can be understood by the fact that an additional ``effective'' pairwise attraction arises between colloids due to the activity \cite{sm}. Even more interesting feature is a macroscopic rotation of the colloid-rich phase, characterized by an angular momentum $\vec{L}$. No such motion is possible in equilibrium. Here $\vec{L}$ is not strictly a constant of motion. Slow random reorientation of the direction of $\vec{L}$ occurs, however, the magnitude of $\vec{L}$ stays almost unchanged.

Coherent motions in dense active matters have been reported for active nematics, mostly in (quasi-) two-dimensional ($2D$) systems. Simulations of active nematics in a quasi-$1D$ microchannel, by Shendruk et al. \cite{shendruk2017}, observed transitions from laminar to oscillatory flow and finally to ``dancing disclinations'', as the strength of the activity is increased. Experiments on confined cellular nematics, such as Retinal cells and mouse myoblasts that form nematic phases, show spontaneous shear flow in monolayers formed on micropatterned glass substrates with adhesive widths in the 100 micrometer range \cite{duclos2018}. Spiral vortical flow was observed in quasi-$2D$ suspensions of Bacillus subtilis in flattened drops with about 50 micrometer diameter \cite{wioland2013}. Simulations of these \cite{woodhouse2012,gao2017} and related flows in annuli and channels  are reviewed in Ref. \cite{doostmohammadi2019}. Studies of $3D$ active fluids in toroidal channels and cylinders (with flows parallel to the cylinder surface) \cite{wu2017} revealed a transition between turbulent motion and coherent flow. The system contained microtubule filaments and depleting polymers (inducing tubuli bundling, i.e., an attraction similar to the AO model) in which activity was generated by the presence of kinesin motor clusters, causing interfilament sliding. Interesting long range effects due to planar boundaries on active matters were also discussed in various contexts \cite{elgeti2013,baek2018,dor2021}. While most of these works deal with strongly elongated objects, we find interesting coherent motion in a system of spherical active particles. 

\section{Model}
While in the original AO model the colloids are hard spheres, for Molecular Dynamics (MD) \cite{allen2017} a smooth repulsion is convenient. Here we use the repulsive part of the Lennard-Jones potential, with strength $\varepsilon=1$ and range $\sigma=1$, the latter taken as the unit of length. The same potential is used for the colloid (c)-polymer (p) interaction as well, but with $\sigma_{cp}=0.9$. The polymers repel each other with a much weaker potential that allows strong polymer-polymer (pp) overlap \cite{zausch2009,sm}, closely resembling the ``no'' pp interaction in the original version. 

All particles are confined in a sphere of radius $R = 10$, the wall potential being the same as in equilibrium studies \cite{winkler2013,statt2012}:
\begin {equation}\label{eq_wall}
U_{wb}(z)=4{\varepsilon}_{wb} \left [\left (\frac{{\sigma}_{wb}}{z} \right )^{12} - \left (\frac{{\sigma}_{wb}}{z} \right )^6+\frac{1}{4}\right ],~b=(c,p).
\end{equation}
Here the interaction strengths $\varepsilon_{wc} = \varepsilon_{wp} = 1$ and $U_{wb}(z)=0$ if the normal distance $z$ from the wall exceeds $2^{1/6}\sigma_{wb}$. Used ranges for the interactions were $\sigma_{wc}=0.6$ and $\sigma_{wp}=0.4$. The simulations solve numerically the Langevin equation, $\vec{r}_n$ being the position of the $n^{\text{th}}$ particle,
\begin{equation}\label{eq_langevin}
m{\ddot{\vec{r}}}_n=-\vec{\nabla}U_n-{\gamma}m{\dot{\vec{r}}}_n+\vec{F}^r_n(t)+{\vec{f}}_n,
\end{equation}
where $m$~$(=1)$ is the mass of each particle. In Eq. (\ref{eq_langevin}), the friction coefficient $\gamma$ is related to the random force ${\vec{F}}^r_n(t)$ via the standard fluctuation-dissipation relation at thermal energy $k_{B}T=1$. Note that $\vec{F}^r_n(t)$ satisfies $\langle \vec{F}^r_n(t) \rangle=0$ and $\langle \vec{F}^r_n(t) \cdot \vec{F}^r_{n'}(t') \rangle=6m{\gamma}k_BT\delta_{nn'}\delta{(t-t')}$. The active force $\vec{f}_n$, that changes only the velocity directions \cite{das2017,chakraborty2020,paul2021}, on particle $n$, is proportional to $f_A{\hat{p}}_n$, where ${\hat{p}}_n$ is an unit vector along the average of the velocities of all particles within a neighborhood of radius $r_{\text{int}}=2^{2/3}$ (see SM for more details).

\begin{figure}[H]
\centering
\includegraphics[scale=1.0]{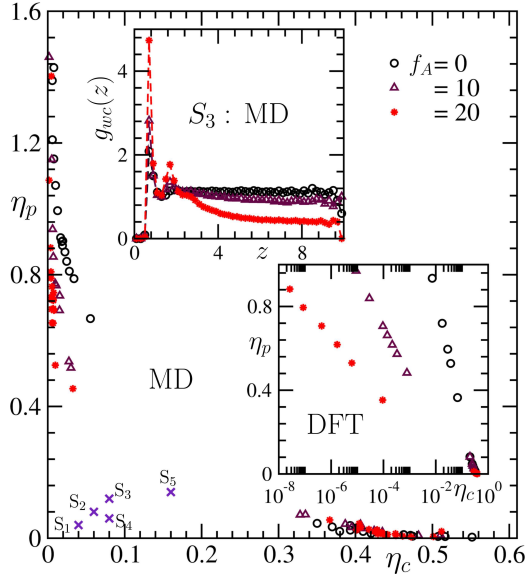}
\caption{Phase diagrams of the active AO model in the $\eta_c$--$\eta_p$ plane, for different strengths of $f_A$ (open circles: $f_A = 0$; triangles: $f_A = 10$; filled circles: $f_A = 20$) of the activity, obtained via MD simulations. Crosses show the state points $S_1,...,S_5$ for which the radial distribution functions, $g_{wb}$, of the particles as function of the radial distance $z$ from the wall were calculated. The upper inset shows the simulation data for, as an example, $g_{wc}(z)$ versus $z$, at $S_3$. Lower inset contains the phase diagrams on a semi-log scale, only for the left branches, as obtained from the density functional theoretical (DFT) calculations (see SM for details), for the same set of $f_A$ values.}
\label{fig1}
\end{figure}

\begin{figure}[H]
\centering
\includegraphics[scale=1.5]{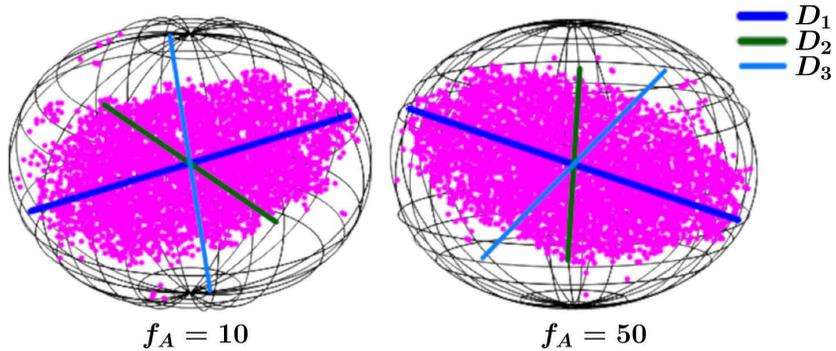}
\caption{Two typical configurations, at time $t=2\times 10^4 \sqrt{m\sigma^2/\varepsilon}$, showing only the polymers inside the sphere, for phase separated states that resulted by fixing the packing fractions at $\eta_c = \eta_p = 0.3$, for  $f_A= 10$ (left)  and $f_A=50$ (right). The principal axes diameters $D_1$, $D_2$ and $D_3$, corresponding to the eigenvalues $\lambda_1 > \lambda_2 > \lambda_3$ of the gyration tensor of the cluster, are indicated by lines.}
\label{fig2}
\end{figure}

\begin{figure}[H]
\centering
\includegraphics[scale=0.5]{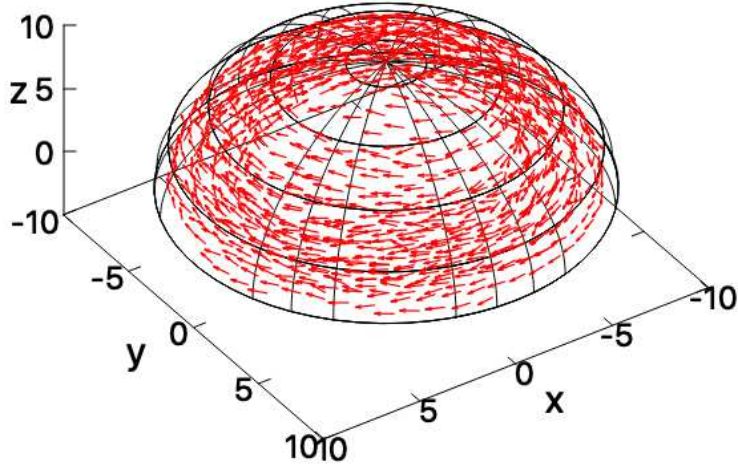}
\caption{Velocity vectors of the colloidal particles inside the sphere. We have bisected the sphere by choosing an equatorial plane that is perpendicular to the eigenvector along $D_1$. Only one hemisphere is shown relative to this plane, by normalizing the velocity magnitudes to unity. The snapshot refers to the same system as in Fig. \ref{fig2}, with $f_A = 50$ at $t = 2 \times 10^4$.}
\label{fig3}
\end{figure}

\begin{figure}[H]
\centering
\includegraphics[scale=0.8]{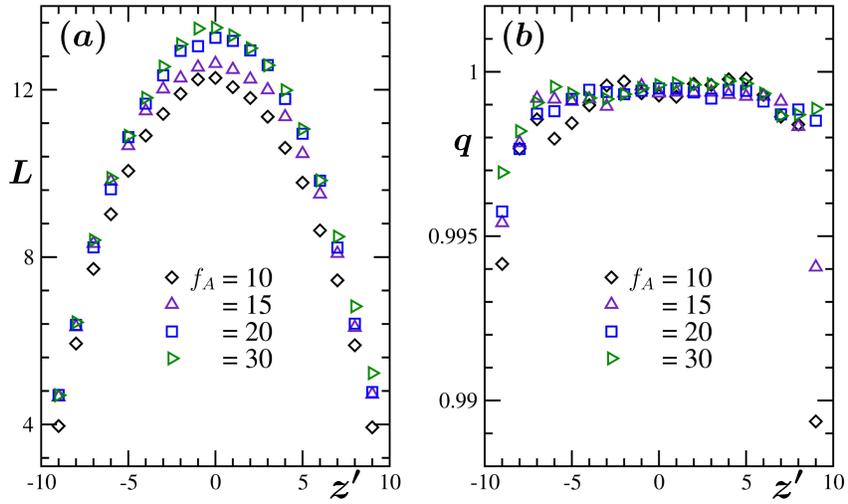}
\caption{For the colloidal particles, the average absolute value of the angular momentum (a) and of the alignment parameter $q$ (b)  are plotted versus the distance $z'$ from the center of the sphere along the diameter $D_1$. Results from different choices of $f_A$ are included, as indicated. All data correspond to the composition $\eta_c = \eta_p= 0.3$.}
\label{fig4}
\end{figure}

\begin{figure}[h]
\centering
\includegraphics[scale=0.8]{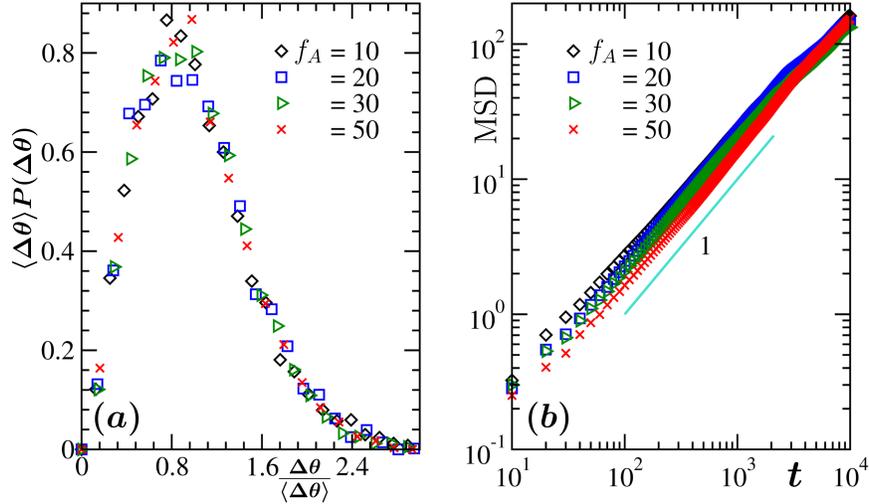}
\caption{(a) Plots showing the scaling collapse of the distribution function $P(\Delta\theta)$ of the angular displacement $\Delta \theta$, over time $t_0=100$, for different values of $f_A$. (b) The mean-squared displacement (MSD) of the poles of the symmetry axis $D_1$ on the sphere is plotted versus time, for different $f_A$. The solid line represents a power-law with the mentioned value of the exponent.}
\label{fig5}
\end{figure}

\section{Results}
The ``phase diagrams'' in the plane of packing fractions $\eta_c$ and $\eta_p$, obtained from the simulations, are shown in the main frame of Fig. \ref{fig1}, for several choices of $f_A$, including the passive case $f_A=0$. States above the symbols are in the immiscible region, the left branches quantifying the polymer-rich and the right ones the colloid-rich phases. The region near the critical point, where the branches are expected to merge, is not displayed here. When $f_A$ increases the phase boundaries move towards the coordinate axes, since activity enhances phase separation. Here $\eta_c$ and $\eta_p$ are related to the colloid and polymer densities $\rho_c$ and $\rho_p$ as \cite{zausch2009} $\eta_c=0.5484\rho_c$ and $\eta_p=0.2808\rho_p$.

The state points $S_1,...,~S_5$, in Fig. \ref{fig1}, were used to study the wall-particle radial distribution functions. As examples, see the upper left inset that shows the results from the state point $S_3$. An increase of $f_A$ favors the surface enrichment of colloids even in the one-phase region (see SM for more details). Treating such data as if they were results for thermal equilibrium, we used density functional theory (DFT) \cite{evans1979,patel2005} to obtain an effective static potential \cite{sm}. From this potential, DFT allows to predict phase diagrams (see the lower right inset of Fig. \ref{fig1}) that are in qualitative agreement with the active matter simulations \cite{sm}. However, as we shall see, closer analysis of the simulation results shows that this is not at all the whole story!

To obtain the simulation data in Fig. \ref{fig1}, runs for a set of state points ($\eta_c$,$\eta_p$) are carried out for large enough $\eta_c$. For each of these, steady state exhibits phase separation with the inner cluster containing mostly polymers, while the colloids accumulate mostly outside this cluster, i.e., near the sphere surface. Fig. \ref{fig2} shows two typical configurations of such polymer clusters. While in the equilibrium case the need to minimize the total interfacial excess free energy should lead to an approximately spherical shape of such a cluster, which actually is observed \cite{winkler2013}, we here find a rather elongated, ellipsoidal cluster shape. Defining the center of mass coordinates $X^i=\sum_{n=1}^{N_p}x_n^i/N_p$ ($i=1,2,3$ in $d = 3$ dimensions and the sum runs over the position coordinates $x_n^i$ of all $N_p$ polymers inside the cluster), the gyration tensor $Q_{ij}$ can be calculated as \cite{solc1971,arkin2013}
\begin{eqnarray}\label{eq_gyration}
    Q_{ij}=\frac{1}{N_p}\sum_{n=1}^{N_p}(x_n^i-X^i)(x_n^j-X^j),~~~i,j=1,2,3.
\end{eqnarray}
Transforming this tensor to principal axes, one finds that the related eigenvalues $\lambda_1 \gg \lambda_2 \simeq \lambda_3$ correspond to a prolate ellipsoidal shape: e.g., $\lambda_1 \simeq 19$ and $\lambda_2 \simeq \lambda_3 \simeq 7.5$ in the example of $f_A = 50$ in Fig. \ref{fig2}. In the steady state the $\lambda_i$s show fluctuations of the magnitude of at most 0.5, without any systematic change with time (see SM).  

Furthermore, the motion of the colloidal particles relative to this structure is interesting. In Fig. \ref{fig3} the snapshot of colloid velocities indicates a coherent collective rotational flow of the particles inside the sphere. This is despite that we have no hydrodynamics in our model. We can quantify this phenomenon in terms of the angular momentum $\vec{L}$, defined as 
\begin{equation}\label{eq_angular}
\vec{L}=\frac{1}{N_{c,z}}\sum_{n=1}^{N_{c,z}} \vec{r}_{\bot c,n} \times \vec{v}_{c,n}.  
\end{equation}
Here $N_{c,z}$ is the total number of colloids within a circular disk at distance $z$, from the cavity centre along the symmetry axis $D_1$, corresponding to $\lambda_1$, with width $dz$, while $\vec{r}_{\bot c,n}$ is the perpendicular distance of the $n^{\text{th}}$ colloid from the rotation axis having unit vector $\hat{e}_1$ corresponding to $\lambda_1$. Fig. \ref{fig4} shows then (a) the average of the magnitude of this angular momentum as well as (b) of an alignment parameter $q = \langle \hat{e}_1 \cdot {\vec{L}}/{|\vec{L}|} \rangle$. Note that $q$ is a measure of how well the axis of this rotation is correlated with the distribution of the polymers in the system. It is seen that in the center of the sphere the magnitude of $\vec{L}$ is maximal and the alignment indeed is nearly perfect. With the same spirit, studying the effective temperatures $T_c$ and $T_p$ of colloids and polymers, from their kinetic energies as in  Ref.  \cite{trefz2016}, we find that $T_p$ is almost uniform but somewhat smaller than the value implied via the noise in Eq. (\ref{eq_langevin}), while $T_c$ is slightly enhanced near the sphere center and depressed near the sphere surface (see SM). 

Finally, we consider the time dependence of the orientation of the eigen vector corresponding to $\lambda_1$, by defining an angle of deviation $\Delta \theta (t_0)$ as
\begin{equation}\label{eq_theta}
\Delta \theta(t_0)=\cos ^{- 1} (|\hat{e}_1(t_0+t) \cdot \hat{e}_1(t)|).
\end{equation}
Fig. \ref{fig5}(a) shows the distribution $P(\Delta\theta)$, for $t_0=100$, demonstrating that it is independent of $f_A$ when $\Delta\theta$ is rescaled with its average. Since the unscaled distribution (see SM) has its peak around $\Delta\theta = 0.1$, the period over which a particle near the equator  takes a full ``roundtrip'' on the sphere is clearly less than $t_0$ (given that $v$ is roughly $2$). Thus, within this time there is only a small change of the orientation of the axis for the rotational flow. This is consistent with Fig. \ref{fig5}(b) where we show that the motion of the endpoint of the rotation axis $\vec{r}_R(t)$ [$=R\hat{e}_1(t)$] is simply diffusive. The diffusion constant has a weak dependence on $f_A$; it decreases only slightly when $f_A$ increases.

\section{Conclusion}
The phase separation caused in colloid-polymer mixtures by depletion forces is enhanced when the colloids are active particles, both in bulk and in spherical confinement. While for equal packing fractions of colloids and polymers, the confining surface remains close to neutral to colloids and polymers in the fully passive system, and, thus, a Janus structure is formed, one observes a different picture in the active case. In this situation, the colloids wet the sphere surface completely, with the polymers forming ellipsoid in the interior. Coherent rotational flow of the colloids occurs around this structure. The principal axis of the ellipsoid exhibits a rotational diffusion, which is slow on the time scale characterizing the rotational flow. Increasing activity slows down the diffusion slightly. In the mixed-phase region of the system, surface enrichment of the colloids occurs, which also is enhanced by the colloid activity. All these findings are qualitatively the same, irrespective of the details of the wall-particle repulsive interactions. It could be speculated that such collective motions in confined active systems are advantageous in the context of biological processes and functions.

\section{Acknowledgements}
The authors are thankful to a National Supercomputing facility at JNCASR. SKD and AB acknowledge partial financial support from SERB, DST, India, via Grant No: MTR/2019/001585. AB thanks CSIR, India, for the research fellowship. 

\section{Supplementary Material}
~~Here we present details of models and methods that were used for simulations and density functional theoretical calculations. We also present some additional results. Unless otherwise mentioned all figures are from simulations. 
\subsection{Simulation Model and Methods}
~~We study the phase behavior and dynamical properties of binary mixtures consisting of active colloids and passive polymers under spherical confinement. In our model, the passive interactions among the particles are taken from a variant of the well-known Asakura-Oosawa (AO) model \cite{asakura1954,asakura1958,vrij1976} of colloid (c) and polymer (p) mixtures. For $\alpha=c$ and $\beta=c,p$, particles $i$ and $j$ interact with each other via the Weeks-Chandler-Andersen (WCA) potential
\begin{equation}\label{inact_AA}
U_{\alpha\beta}=
\begin{cases}
4{\varepsilon}_{\alpha \beta} \left [\left (\frac{{\sigma}_{\alpha \beta}}{r} \right )^{12} - \left (\frac{{\sigma}_{\alpha \beta}}{r} \right )^6+\frac{1}{4}\right ], & \text{for}~r<2^{\frac{1}{6}}{\sigma}_{\alpha \beta}\\
0, & \text{otherwise;}\\
\end{cases}
\end{equation}
and we have for pp pairs
\begin {equation}\label{inact_BB}
U_{pp}(r)=
\begin{cases}
8{\varepsilon}_{pp} \left [1-10\left (\frac{r}{r_{c,pp}} \right )^{3} + 15\left (\frac{r}{r_{c,pp}} \right )^4-6\left (\frac{r}{r_{c,pp}} \right )^5\right ], & \text{for}~r<r_{c,pp}=2^{\frac{1}{6}}{\sigma}_{pp}\\
0, & \text{otherwise.}\\
\end{cases}
\end{equation}
Here $\varepsilon_{\alpha\beta}$ and $\sigma_{\alpha\beta}$ are the interaction strengths and interaction diameters for various combinations of species. We set \cite{das2014,trefz2016} ${\varepsilon}_{cc}={\varepsilon}_{cp}={\varepsilon}=1$, ${\varepsilon}_{pp}=0.0625$, ${\sigma}_{cc}={\sigma}=1$, ${\sigma}_{cp}=0.9$ and ${\sigma}_{pp}=0.8$. 
\par
~~The wall-particle interactions are modeled by smooth repulsions described by the WCA potential
\begin {equation}\label{wall_par}
U_{wb}(z)=
\begin{cases}
4{\varepsilon}_{wb} \left [\left (\frac{{\sigma}_{wb}}{z} \right )^{12} - \left (\frac{{\sigma}_{wb}}{z} \right )^6+\frac{1}{4}\right ], & \text{for }z<2^{\frac{1}{6}}{\sigma}_{wb}\\
0, & \text{else.}
\end{cases}
\end{equation}
Here $b=(c,p)$ and $z$ is the shortest distance between the particle and the spherical wall. The interaction strength between the wall and the particles (c or p) are taken to be unity, i.e., ${\varepsilon}_{wc}={\varepsilon}_{wp}=1$. The relative influence of the wall on the particles is controlled by the ratio of the repulsive ranges \cite{winkler2013}, viz., $q_w={\sigma}_{wp}/{\sigma}_{wc}$, where $\sigma_{wc}$ and $\sigma_{wp}$ are the diameters for wall-colloid and wall-polymer interactions, respectively. 
\par
~~Molecular dynamics (MD) simulations have been performed via the numerical solutions of the Langevin equation ($\vec{r}_n$ is the position of $n^{\text{th}}$ particle)
\begin{equation}
m\ddot{\vec{r}}_n=-\vec{\nabla}U_n-{\gamma}m\dot{\vec{r}}_n+\vec{F}^r_n(t)+{\vec{f}}_n.
\end{equation}
Here the mass $m$ is same for all colloid and polymer particles; $\gamma$ is a damping coefficient; $\vec{F}^r_n$ is a random force; and $\vec{f}_n$ is the active force. The random force is ${\delta}$-correlated as
\begin{equation}
\langle \vec{F}^r_n(t) \cdot \vec{F}^r_{n'}(t') \rangle=6m{\gamma}k_BT\delta_{nn'}\delta{(t-t')}.
\end{equation}
\par
~~We have ${\vec{f}}_n=0$ for all polymers and the self-propelling activity among the colloids is introduced via the Vicsek model in which the velocity of the $n^{\text{th}}$ colloid ($\vec{v}_{c,n}$) gets influenced by the average direction ${\hat{p}}_n$ ($={{\sum_k}{\vec{v}_{c,k}}}/{\lvert {\sum_k}{\vec{v}_{c,k}} \rvert}$) of colloids within its neighbourhood (including $n^{\text{th}}$ colloid) that is defined by a distance $r_{\text{int}}$ from $\vec{r}_n$. The active force, $\vec{f}_n$, on the $n^{\text{th}}$ colloid is proportional to $f_A{\hat{p}}_n$, where $f_A$ is the strength of activity. The activity has been incorporated in such a way that only the direction of velocity changes, not the magnitude. This way of implementation of activity will allow the system to be in the assigned temperature \cite{das2017}. 
\par
~~The mass, length and time in our simulations were measured in units of $m$, $\sigma$ and $t_0=\sqrt{{\sigma^2}m/\varepsilon}$, respectively. The time step in the MD simulations is taken to be ${\Delta}t=0.002t_0$. We have fixed $m$, $\sigma$, $\varepsilon$, $k_B$, $T$ and $\gamma$ to unity and used $r_{\text{int}}=2^{2/3}$. All the results are presented for $R=10$, the radius of the spherical cavity. In this work we use $\sigma_{wc}=0.6$ and $\sigma_{wp}=0.4$, unless otherwise mentioned. For this choice of $q_w$ (=0.667), there is no preferential affinity of wall towards any components, in the full passive case ($f_A=0$). In equilibrium, in this case, both components wet the wall forming Janus particle-like structure under conditions where the system exhibits phase separation in the bulk. Introduction of activity to the colloids leads to drastic change in configurations in steady states with respect to passive configurations in equilibrium. We have chosen these particular values for the parameters ${\sigma}_{wb}$ for certain reasons. Note that the wetting properties of this model in the passive limit have been studied by A. Statt et. al. \cite{statt2012}. Complete wetting of the sphere surface by the colloids was observed for up to ${\sigma}_{wc}=0.59$ and partial wetting picture emerged for ${\sigma}_{wc}$ lying approximatly between 0.59 and 0.7, for ${\sigma}_{wp}=0.4$. The case of ``neutral'' walls, i.e., a contact angle of $90^0$, was found for ${\sigma}_{wc}=0.65$. All data are for packing fractions $\eta_c=0.3$ and $\eta_p=0.3$ of both polymers and colloids, unless otherwise mentioned. 
\subsection{Selected Additional Simulation Results}
\begin{figure}[H]
\centering
\includegraphics[scale=0.8]{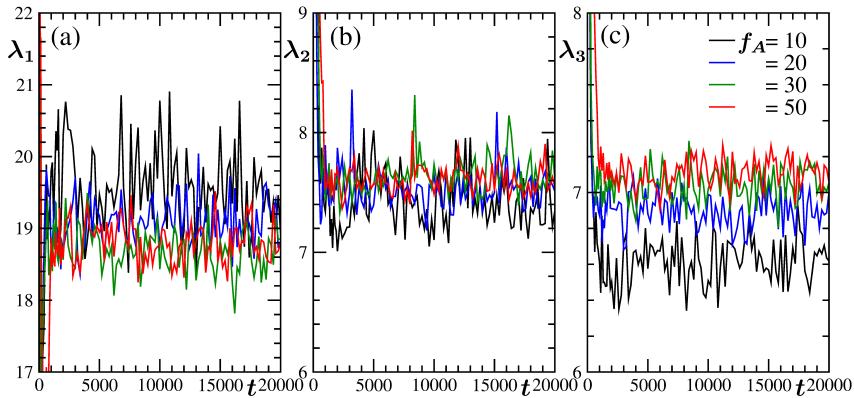}
\caption{(a)-(c) The eigenvalues $\lambda_i$ ($i=1,2,3$) have been plotted, for different values of the activity strength $f_A$, as a function of time. These results are for $\sigma_{wc}=0.6$ and $\sigma_{wp}=0.4$.} 
\label{fig_s1}
\end{figure}

From the gyration tensor $Q_{ij}$ ($i,j=1,2,3$) (see the main article) of all the polymers within the cavity, we have calculated the eigenvalues $\lambda_i$~($i=1,2,3$) during the evolution of a system. In Fig. \ref{fig_s1}(a)-(c) we have shown the change of these eigenvalues, versus time, for different $f_A$. We have divided the sphere into many circular disks, each of width $dz$, along the principal axis diameter $D_1$, corresponding to $\lambda_1$, of the ellipsoid that is formed by the polymers in the interior of the sphere. Let us consider a disk $C_z$, situated at a distance $z$ along $D_1$ from the centre of the cavity, having $N_{c,z}$ colloids and and $N_{p,z}$ polymers.

\begin{figure}[H]
\centering
\includegraphics[scale=0.8]{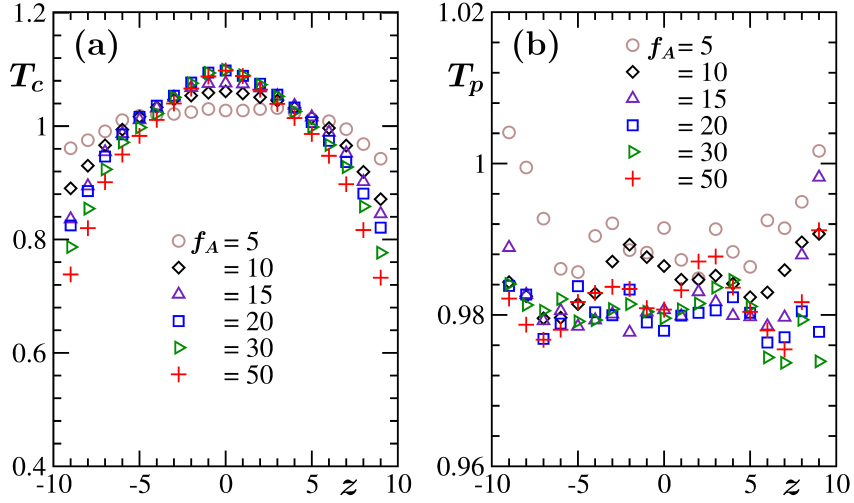}
\caption{The average temperature of colloids, $T_c$, in (a), and of polymers, $T_p$, in (b), are plotted versus the distance $z$ along the diameter $D_1$. Data for several $f_A$ have been shown. All the presented results are for $\sigma_{wc}=0.6$ and $\sigma_{wp}=0.4$.	} 
\label{fig_s2}
\end{figure}

The average temperatures of colloids ($T_c$) and polymers ($T_p$) belonging to the disk $C_z$ are calculated as
\begin{equation}
\begin{split}
T_c(z) & =\frac{1}{3k_B N_{c,z}}\sum_{k \in C_z} {{\vec{v}_{c,k}}}^2,\\
T_p(z) & =\frac{1}{3k_B N_{p,z}}\sum_{k \in C_z} {{\vec{v}_{p,k}}}^2.
\end{split}
\end{equation}
The sum is carried over the particles ($c$ or $p$) within the disk $C_z$. Here $\vec{v}_{c,k}$ and $\vec{v}_{p,k}$ are the velocities of the $k^{\text{th}}$ colloid and polymer, respectively. We have shown the variation of $T_c$ and $T_p$ with $z$, for different $f_A$ , in Fig. \ref{fig_s2}(a)-(b).    

\begin{figure}[H]
\centering
\includegraphics[scale=0.8]{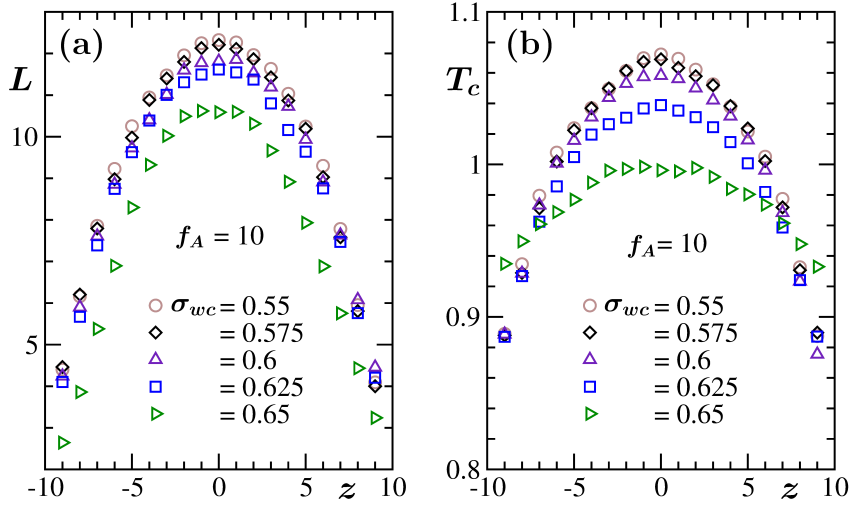}
\caption{(a) Plots of magnitude of the average of the angular momentum, $L$, versus $z$, have been shown for different $\sigma_{wc}$. (b) Average temperatures of colloid ($T_c$) are plotted versus $z$, for different $\sigma_{wc}$. The results in both the frames are obtained for $\sigma_{wp}=0.4$ and $f_A=10$.}  
\label{fig_s3}
\end{figure}

The average angular momentum $\vec{L}(z)$, with respect to the diameter $D_1$, of the colloids inside ${C}_z$, is defined as 
\begin{equation}
\vec{L}(z)=\frac{1}{N_{c,z}}\sum_{k\in {C}_z} \vec{r}_{\bot c,k} \times \vec{v}_{c,k}.
\end{equation}
Here $\vec{r}_{\bot c,k}$ is the perpendicular distance of the $k^{\text{th}}$ colloid particle from $D_1$: $\vec{r}_{\bot c,k}=\vec{r}_{c,k}-\hat{e}_1(\hat{e}_1 \cdot \vec{r}_{c,k})$, $\hat{e}_1$ being the unit vector along $D_1$. Note that $\vec{r}_{c,k}$ is the position vector of the $k^{\text{th}}$ colloid. The variations of $L~(=|\vec{L}|)$ and $T_c$ with $z$ are presented in Fig. \ref{fig_s3}(a)-(b) for different wall-colloid interaction ranges $\sigma_{wc}$, by setting $\sigma_{wp}=0.4$ and $f_A=10$. Below we define a few average values that may be useful in quantifying the whirling motion. 

\begin{figure}[H]
\centering
\includegraphics[scale=0.8]{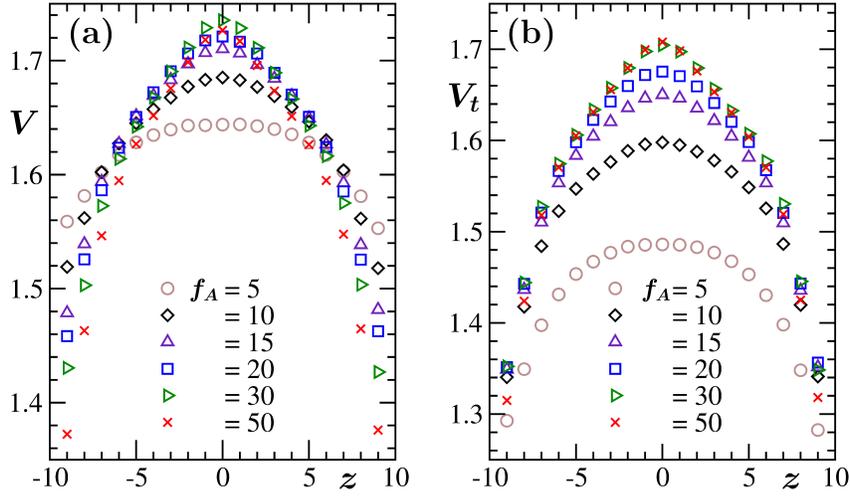}
\caption{The average of the magnitudes of the velocities of the colloids, as a function of $z$, are shown in (a) and the average of the tangential velocity components with respect to $D_1$ are plotted versus $z$ in (b), for different $f_A$. These results were obtained by fixing $\sigma_{wc}$ at 0.6 and $\sigma_{wp}$ at 0.4.}  
\label{fig_s4}
\end{figure}

The average magnitude of the velocities of colloids ($V$) inside the disk $C_z$ is calculated as
\begin{equation}
V(z)=\frac{1}{N_{c,z}}\sum_{k \in C_z} |\vec{v}_{c,k}|.  
\end{equation}
For the tangential component of the velocities of colloids around $D_1$ in the disk $C_z$, the average value ($V_t$) is calculated as
\begin{equation}
V_t(z)=\frac{1}{N_{c,z}}\sum_{k \in C_z} |\vec{v}_{c,k}-\hat{r}_{\bot c,k}(\vec{v}_{c,k} \cdot \hat{r}_{\bot c,k})|,  
\end{equation}
where $\hat{r}_{\bot c,k}=\frac{\vec{r}_{\bot c,k}}{|\vec{r}_{\bot c,k}|}$. The average perpendicular distance ($r_p$) of the colloids, situated inside $C_z$, from the axis corresponding to $D_1$, is given by
\begin{equation}
r_p(z)=\frac{1}{N_{c,z}}\sum_{k \in C_z} |\vec{r}_{\bot c,k}|.  
\end{equation}
The average value of time ($t_r$) taken by colloids to complete one rotation around $D_1$ is $t_r=2{\pi}r_p/V_t$. 

In Fig. {\ref{fig_s4}}(a) and (b) we have shown the variation of the average absolute velocity of colloids $V$ and the average tangential component (of velocity) $V_t$ with the change of $z$, for different $f_A$. 

\begin{figure}[H]
\centering
\includegraphics[scale=0.8]{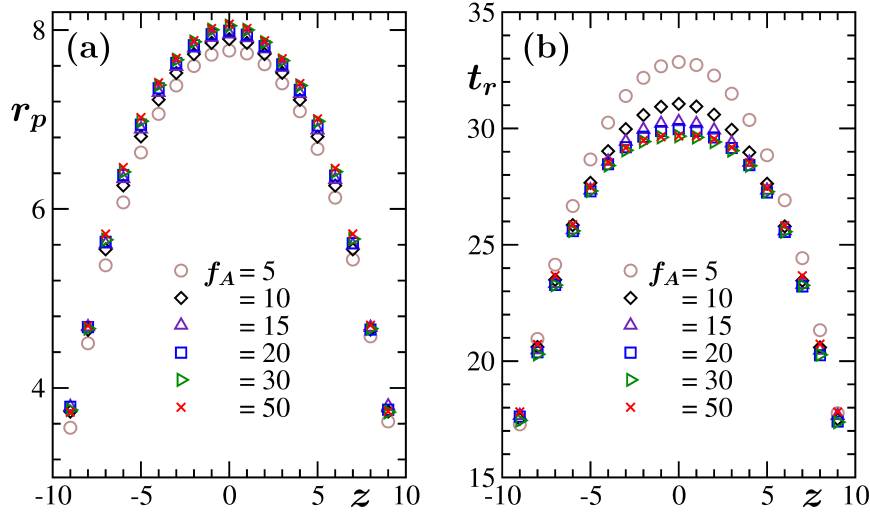}
\caption{(a) The average perpendicular distance of the colloids from $D_1$ is shown as a function of $z$. Data for several choices of $f_A$ are shown. (b) The average time $t_r$ ($=2{\pi}r_p/V_t$) for one complete rotation around $D_1$ by colloids are shown for different $f_A$, versus $z$. The presented data sets correspond to $\sigma_{wc}=0.6$ and $\sigma_{wp}=0.4$.}  
\label{fig_s5}
\end{figure}

The average perpendicular distance $r_p$ of colloids from $D_1$ and the average time $t_r$ for one complete rotation of colloids around $D_1$ are plotted versus $z$ in Fig. \ref{fig_s5}(a) and (b) for different $f_A$.

\begin{figure}[H]
\centering
\includegraphics[scale=0.5]{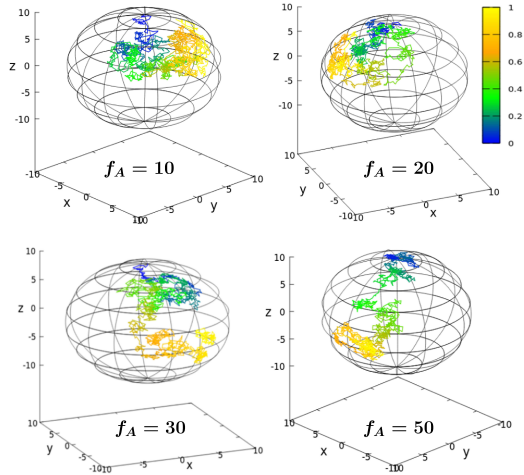} 
\caption{We have shown the trajectories of chosen end points of the symmetry diameters $D_1$ for different $f_A$. The color code represents the time evolution of a trajectory in steady state of a system in a manner described below. The coordinate systems are transformed in such a way that the starting point is the north pole. The trajectory is shown from a starting time $t_{\text{min}}$~($=10^4$) to a maximum time $t_{\text{max}}$~($=2.5 \times 10^4$). We set the intensity of the color bar according to $q_{cb}=\frac{t-t_{\text{min}}}{t_{\text{max}}-t_{\text{min}}}$. The presented results  are for $\sigma_{wc}=0.6$ and $\sigma_{wp}=0.4$.}  
\label{fig_s6}
\end{figure}

In Fig. \ref{fig_s6} we have shown a few typical trajectories of chosen end points of the symmetry diameters $D_1$. The trajectories lie on the surface of the sphere. For these the mean-squared displacement (MSD) is calculated as
\begin{equation}
\text{MSD}(t)=\langle {(\Vec{r}_i(t)-(\Vec{r}_i(0))}^2 \rangle,
\end{equation}
where $\vec{r}_i(t)=R\hat{e}_1(t)$, with the average being obtained by considering many such trajectories, for each of the $f_A$ values.

\begin{figure}[H]
\centering
\includegraphics[scale=0.4]{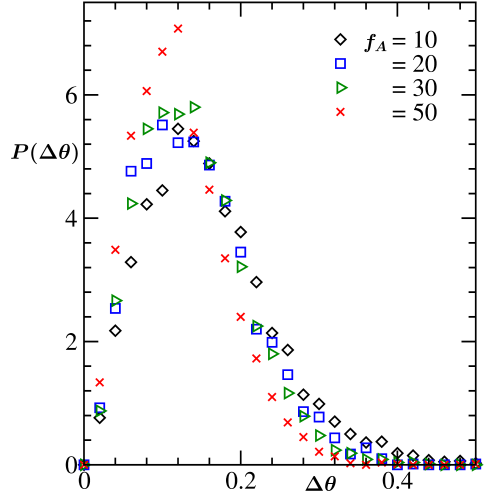}
\caption{The distributions of the angular displacement $\Delta \theta$ are shown for different values of $f_A$. $P(\Delta \theta)$ in each of the cases has been obtained by fixing $t_0$ at $100$. The observation times ($t$) belong to the range [$10^4, 3 \times 10^4$]. The distribution, for each $f_A$, is obtained from a total of 4000 $\Delta \theta$ values (collected from different runs) lying in the above mentioned time range. These results are for $\sigma_{wc}=0.6$ and $\sigma_{wp}=0.4$.}  
\label{fig_s7}
\end{figure}

The angular displacement $\Delta \theta$ of $D_1$ over time $t_0$ between two steady states at times $t$ and $t+t_0$ is defined as
\begin{equation}
\Delta \theta(t_0)={\cos}^{- 1} (|\hat{e}_1(t+t_0) \cdot \hat{e}_1(t)|).
\end{equation}
We have estimated the distributions of $\Delta \theta$ for different $f_A$. The probability density functions $P(\Delta \theta)$ have been obtained by fixing $t_0$ at $100$ and are shown in Fig. \ref{fig_s7}.


\subsection{Details of Density Fuctional Theory} 
\label{section1}

\subsubsection{Microscopic Model: Passive Colloids}
\label{subsection1.1}

There are some minor differences between the microscopic models employed in MD simulations and in our Density Functional Theory (DFT) calculations. In particular, all the pairwise interactions in the MD-studied model are represented by continuous interaction potentials: the repulsive part of the Lennard-Jones (LJ) potential for colloid-colloid and colloid-polymer pair interactions, and a weakly repulsive potential for polymer-polymer interactions. By contrast, as discussed in detail below, the DFT-studied model uses hard-sphere potentials to model colloid-colloid and colloid-polymer pair interactions, while polymer-polymer interactions are taken to be ideal, following the standard Asakura-Oosawa (AO) model~\cite{asakura1954,asakura1958,vrij1976}. The reason for these differences stems from the fact that MD simulations are easier to perform with continuous potentials, while DFT formalism has been developed and tested~\cite{egorov2021} for the standard AO model, where highly accurate excess free energy functional based on the Fundamental Measure Theory~\cite{roth2010} has been constructed~\cite{schmidt2002}.

Thus, for a passive binary colloid-polymer mixture we consider the standard AO model~\cite{asakura1954,asakura1958,vrij1976} in a spherical cavity of radius $R=10$ with volume $V=4\pi R^3/3$. Colloidal particles are represented by $N_c$ hard spheres of diameter $\sigma_c$, while ideal polymers are represented by $N_p$ noninteracting spheres of diameter $\sigma_p$ ($\sigma_c=1$ will be taken as the unit of length throughout). The size ratio of polymers and colloids, $q_s=\sigma_p/\sigma_c$, controls the range of the depletion attraction between the colloidal particles~\cite{asakura1954,asakura1958,vrij1976}. It will be kept fixed at the value of $q_s=0.8$ in this study. The packing fractions of colloids and polymers are given by $\eta_b=(\pi\sigma_{b}^{3}N_{b})/(6V)$, where $b=c,p$ for colloids and polymers, respectively.

The pairwise interparticle interactions in the AO model are described by the following set of spherically symmetric potentials~\cite{asakura1954,vrij1976,asakura1958}:
\begin{equation}
  v_{cc}(R_{ij})=\infty, \quad R_{ij}<\sigma_c; \quad v_{cc}(R_{ij})=0, \quad R_{ij}>\sigma_c,
  \label{vcc}
\end{equation}
for two colloidal particles with centers of mass located at ${\bf{R_i}}$ and ${\bf{R_j}}$, separated by the distance 
$R_{ij}=|{\bf{R_i}}-{\bf{R_j}}|$;
\begin{equation}
  v_{pp}(r_{ij})=0,
  \label{vpp}
\end{equation}
for two ideal polymers  with centers of mass located at ${\bf{r_i}}$ and ${\bf{r_j}}$, separated by the distance 
$r_{ij}=|{\bf{r_i}}-{\bf{r_j}}|$; and
\begin{equation}
  v_{cp}(|{\bf{R_i}}-{\bf{r_j}}|)=\infty, \quad |{\bf{R_i}}-{\bf{r_j}}|<(\sigma_c+\sigma_p)/2, \quad v_{cp}(|{\bf{R_i}}-{\bf{r_j}}|)=0, \quad |{\bf{R_i}}-{\bf{r_j}}|>(\sigma_c+\sigma_p)/2,
  \label{vcp}
\end{equation}
for the colloid-polymer pair.

To model the interactions of both colloids and polymers with the confining spherical cavity walls, we use hard-core repulsive potentials $v_{wc}(r)$ and $v_{wp}(r)$, respectively~\cite{statt2012,winkler2013}:
\begin{equation}
  v_{wc}(r)=\infty, \quad r<\sigma_{wc}, \quad v_{wc}(r)=0, \quad r>\sigma_{wc},
  \label{vcw}
\end{equation}
and 
\begin{equation}
  v_{wp}(r)=\infty, \quad r<\sigma_{wp}, \quad v_{wp}(r)=0, \quad r>\sigma_{wp},
  \label{vpw}
\end{equation}
where $r$ is the normal distance between the particle's center of mass and the confining spherical cavity wall. In this work we keep the wall-polymer interaction range fixed at $\sigma_{wp}=0.4$, while the (passive) wall-colloid interaction range is fixed at $\sigma_{wc}=0.6$~\cite{statt2012,winkler2013}. 

\subsubsection{Microscopic Model: Active Colloids}
\label{subsection1.2}

Following our earlier work~\cite{das2014,trefz2016} on active colloid-passive polymer mixtures in the bulk, we extend this work to the case of spherical confinement by mapping the confined active mixtures onto effective passive systems. This approach involves modifying both particle-particle and wall-particle interaction potentials given in Section~\ref{subsection1.1} by adding a soft effective component arising due to the colloid activity. In particular the total effective colloid-colloid interaction potential now takes the form:
\begin{equation}
  v_{cc}^{\rm eff}(R_{ij})=v_{cc}(R_{ij})+\phi_{cc}(R_{ij}),
  \label{vcceff}
\end{equation}
where the soft effective component $\phi_{cc}(R_{ij})$ is obtained from the MD simulation data for the corresponding {\em active} system via the Boltzmann iterative inversion procedure as described in detail in our earlier work~\cite{das2014,trefz2016}.

Likewise, the total effective wall-colloid interaction potential now takes the form:
\begin{equation}
  v_{wc}^{\rm eff}(r)=v_{wc}(r)+\phi_{wc}(r),
  \label{vcweff}
\end{equation}
where the inversion procedure is now applied to the MD-simulated wall-colloid distribution function of the corresponding active system, in order to obtain the soft effective component $\phi_{wc}(r)$. 

\subsubsection{Density Functional Theory: Passive Colloids}
\label{subsection1.3}

The starting point of the DFT approach~\cite{evans1979,evans1992,schmidt2002,egorov2004} is the expression for the grand-canonical free energy $\Omega[\rho_{c}({\bf{r}}),\rho_{p}({\bf{r}})]$, which is written as a functional of one-particle distribution functions $\rho_{b}({\bf{r}})$ ($b=c,p$)~\cite{schmidt2002}:
\begin{equation}
  \beta\Omega[\rho_{c}({\bf{r}}),\rho_{p}({\bf{r}})]=
  \sum_{b=c,p}\int d{\bf{r}}\rho_{b}({\bf{r}})[\ln(\rho_{b}({\bf{r}})\lambda_{b}^{3})-1]
  +\beta F_{\rm exc}[\rho_{c}({\bf{r}}),\rho_{p}({\bf{r}})]+
  \beta\sum_{b=c,p}\int d{\bf{r}}\rho_{b}({\bf{r}})[v_{wb}({\bf{r}})-\mu_b],
\label{omega}
\end{equation}
where $\beta=1/(k_BT)$, $\lambda_{b}$ is the de Broglie wavelength of species $b$, $\mu_b$ is the chemical potential of species $b$, and $v_{wb}$ is the external potential acting on species $b$, as given by Eqs.~(\ref{vcw}) and (\ref{vpw}). The excess Helmholtz free energy functional $F_{\rm exc}[\rho_{c}({\bf{r}}),\rho_{p}({\bf{r}})]$ is written as an integral over excess free energy density $\Phi$, which is a function of weighted densities~\cite{schmidt2002}:
\begin{equation}
  \beta F_{\rm exc}[\rho_{c}({\bf{r}}),\rho_{p}({\bf{r}})]=
  \int d{\bf{r}}^{\prime}\Phi(\{n_{c}^{(\alpha)}({\bf{r}}^{\prime})\},\{n_{p}^{(\beta)}({\bf{r}}^{\prime})\}),
\label{fexc}  
\end{equation}
where the weighted densities for both species are given by convolutions of their respective density profiles with the corresponding weight functions:
\begin{equation}
  n_{b}^{(\alpha)}({\bf{r}})=\int d{\bf{r}}^{\prime}\rho_{b}({\bf{r}}^{\prime})
  \omega_{b}^{(\alpha)}({\bf{r}}-{\bf{r}}^{\prime})
  \label{nscalar}
\end{equation}  
and
\begin{equation}
  {\bf{n}}_{b}^{(\alpha)}({\bf{r}})=\int d{\bf{r}}^{\prime}\rho_{b}({\bf{r}}^{\prime})
  {\bm\omega}_{b}^{(\alpha)}({\bf{r}}-{\bf{r}}^{\prime}),
  \label{nvector}
\end{equation}  
where $n_{b}^{(\alpha)}({\bf{r}})$ and ${\bf{n}}_{b}^{(\alpha)}({\bf{r}})$ denote the scalar and vector weighted densities, respectively. 

The four scalar weight functions are given by~\cite{schmidt2002}:
\begin{equation}
  \omega_{b}^{(3)}({\bf{r}})=\Theta(R_b-r),
  \label{omega3}
\end{equation}  
\begin{equation}
  \omega_{b}^{(2)}({\bf{r}})=\delta(R_b-r),
  \label{omega2}
\end{equation}  
\begin{equation}
  \omega_{b}^{(1)}({\bf{r}})=\frac{\omega_{b}^{(2)}({\bf{r}})}{4\pi R_b},
  \label{omega1}
\end{equation}  
and
\begin{equation}
  \omega_{b}^{(0)}({\bf{r}})=\frac{\omega_{b}^{(2)}({\bf{r}})}{4\pi R_{b}^{2}},
  \label{omega0}
\end{equation}  
where $\Theta(r)$ is the Heaviside step function, $\delta(r)$ is the Dirac delta function, and
$R_b=\sigma_b/2$ are the radii of colloids and polymers. 

The two vector weight functions are given by~\cite{schmidt2002}:
\begin{equation}
  {\bm{\omega}}_{b}^{(2)}({\bf{r}})=\frac{{\bf{r}}}{r}\delta(R_b-r),
  \label{omegav2}
\end{equation}  
and
\begin{equation}
  {\bm{\omega}}_{b}^{(1)}({\bf{r}})=\frac{{\bm{\omega}}_{b}^{(2)}({\bf{r}})}{4\pi R_b}.
  \label{omegav1}
\end{equation}  

The excess free energy density $\Phi_{\rm exc}$ can be written as a sum of three terms~\cite{schmidt2002}:
\begin{equation}
  \Phi_{\rm exc}=\Phi_1+\Phi_2+\Phi_3,
  \label{phiexc}
\end{equation}  
where
\begin{equation}
  \Phi_1=n_{c}^{(0)}[-\ln(1-n_{c}^{(3)})+\frac{n_{p}^{(3)}}{1-n_{c}^{(3)}}]-n_{p}^{(0)}\ln(1-n_{c}^{(3)}),
  \label{phiexc1}
\end{equation}  

\begin{equation}
  \Phi_2=(n_{c}^{(1)}n_{c}^{(2)}-{\bf{n}}_{c}^{(1)}\cdot{\bf{n}}_{c}^{(2)})[\frac{1}{1-n_{c}^{(3)}}+
    \frac{n_{p}^{(3)}}{(1-n_{c}^{(3)})^2}]+
  \frac{n_{p}^{(1)}n_{c}^{(2)}-{\bf{n}}_{p}^{(1)}\cdot{\bf{n}}_{c}^{(2)}+n_{c}^{(1)}n_{p}^{(2)}-
    {\bf{n}}_{c}^{(1)}\cdot{\bf{n}}_{p}^{(2)}}{1-n_{c}^{(3)}},
  \label{phiexc2}
\end{equation}  
and
\begin{equation}
  \Phi_3=\frac{(n_{c}^{(2)})^3-3n_{c}^{(2)}{\bf{n}}_{c}^{(2)}\cdot{\bf{n}}_{c}^{(2)}}{24\pi}
      [\frac{1}{(1-n_{c}^{(3)})^2}+
    \frac{2n_{p}^{(3)}}{(1-n_{c}^{(3)})^3}]+
      \frac{n_{p}^{(2)}(n_{c}^{(2)})^2-n_{p}^{(2)}{\bf{n}}_{c}^{(2)}\cdot{\bf{n}}_{c}^{(2)}
        -2n_{c}^{(2)}{\bf{n}}_{c}^{(2)}\cdot{\bf{n}}_{p}^{(2)}}{8\pi(1-n_{c}^{(3)})^2}.
  \label{phiexc3}
\end{equation}  

The ideal free energy is given by~\cite{schmidt2002}:
\begin{equation}
\beta F_{\rm bid}=\sum_{b=c,p}\int d{\bf{r}}\rho_{b}({\bf{r}})[\ln(\rho_{b}({\bf{r}})\lambda_{b}^{3})-1].
  \label{fideal}
\end{equation}
 
By minimizing $\beta\Omega[\rho_{c}({\bf{r}}),\rho_{p}({\bf{r}})]$ given in Eq.~(\ref{omega}), one obtains the equilibrium inhomogeneous density profiles of colloids and polymers in confined systems described by the external potentials given by Eqs.~(\ref{vcw}) and (\ref{vpw}). The numerical procedure for performing this calculation is detailed in Section~\ref{subsection1.5}. 

\subsubsection{Density Functional Theory: Active Colloids}
\label{subsection1.4}

As discussed in Section~\ref{subsection1.2}, the major difference in the microscopic models of passive and active colloid-polymer mixtures, is the replacement of colloid-colloid and wall-colloid interaction potentials given by Eqs.~(\ref{vcc}) and (\ref{vcw}), respectively, by the corresponding effective interaction potentials given by Eqs.~(\ref{vcceff}) and (\ref{vcweff}), respectively. The latter potentials are constructed by applying the iterative inversion procedure~\cite{das2014,trefz2016} to the simulated structural properties of the active mixture (colloid-colloid and wall-colloid distribution functions). As a result, $v_{wc}(r)$ in the last term of Eq.~(\ref{omega}) gets replaced with $v_{wc}^{\rm eff}(r)$, while the excess Helmholtz free energy functional (the second term in Eq.~(\ref{omega})) acquires an additional contribution due to the soft effective colloid-colloid interaction:
\begin{equation}
  \beta F_{\rm exc}[\rho_{c}({\bf{r}}),\rho_{p}({\bf{r}})]=
  \int d{\bf{r}}^{\prime}\Phi(\{n_{c}^{(\alpha)}({\bf{r}}^{\prime})\},\{n_{p}^{(\beta)}({\bf{r}}^{\prime})\})+
\beta F_{\rm exc}^{\rm eff}[\rho_{c}({\bf{r}})].
  \label{fexcactive}  
\end{equation}
Keeping in mind that the soft effective component $\phi_{cc}(r)$ is a relatively slowly varying function of the colloid-colloid separation, this additional term is computed using the standard mean-field approximation~\cite{patel2005,loverso2012}: 
\begin{equation}
\beta F_{\rm exc}^{\rm eff}[\rho_{c}({\bf{r}})]=\frac{\beta}{2}\int d{\bf{r}}\int d{\bf{r}}^{\prime}\rho_{c}({\bf{r}})\rho_{c}({\bf{r}}^{\prime})\phi_{cc}(|{\bf{r}}-{\bf{r}}^{\prime}|).
\label{fatt}
\end{equation}

\subsubsection{Numerical Implementation}
\label{subsection1.5}

In order to minimize $\beta\Omega[\rho_{c}({\bf{r}}),\rho_{p}({\bf{r}})]$, we use a straightforward Picard iterative procedure~\cite{roth2010}, which was found to be adequate for the present simple microscopic model~\cite{egorov2021}. 
As initial guess for the density profiles, we employ the corresponding ideal gas solutions: 
$\rho_{b}(r)=\rho_{b}^{0}\exp(-\beta v_{wb}(r))$, $b=c,p$. The resulting equilibrium density profiles have the same spatial symmetry as the external potentials $v_{wb}(r)$, i.e. the inhomogeneous density profiles $\rho_b(r)$ are functions of a single variable, radial distance $r$ from the confining spherical wall. For the spherical geometry, the integration over the remaining two spherical polar coordinates can be performed analytically. Hence, the three-dimensional convolutions in Eqs.~(\ref{nscalar}) and (\ref{nvector}) reduce to one-dimensional integrations~\cite{roth2010,egorov2021}. The latter are performed on an equidistant radial grid with the spacing of $dr=0.02$ using 2-point Gaussian quadrature. The tolerance criterion for terminating the iterative numerical procedure is set to $10^{-6}$. 

\begin{figure}
\includegraphics[scale=0.4]{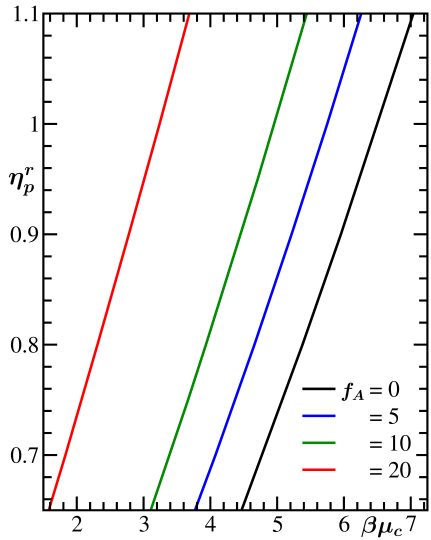} 
\caption{Phase diagram for a spherical cavity of radius $R=10$ in the ($\eta_{p}^{r}$, $\beta\mu_c$) plane; the model parameters are fixed at $\sigma_{wc}$=0.6, $\sigma_{wp}=0.4$ and $q_s=0.8$. The results are presented for four values of the activity strength $f_A$, including the passive mixture, $f_A=0$. These results are obtained via DFT calculations.} 
\label{fig_s8}
\end{figure}

\begin{figure}
\includegraphics[scale=0.8, angle=0]{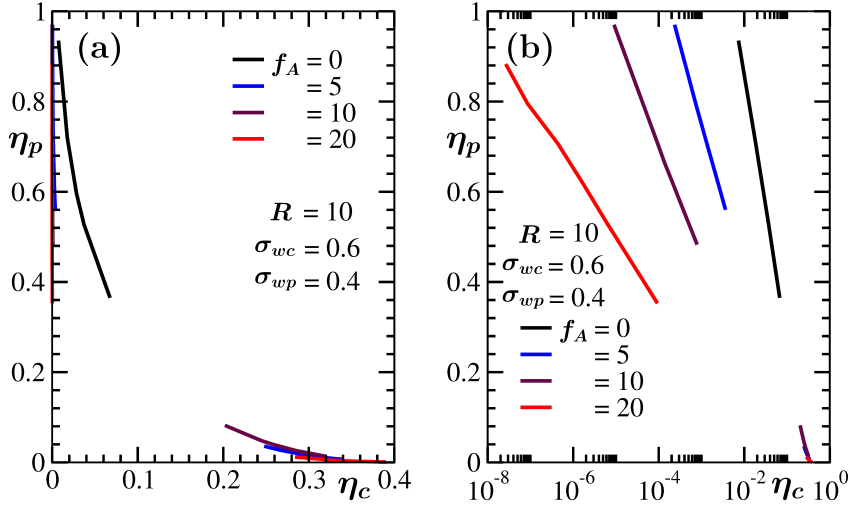}
\caption{(a) Phase diagram for a spherical cavity of radius $R=10$ in the ($\eta_{p}$, $\eta_{c}$) plane; the model parameters are fixed at $\sigma_{wc}$=0.6, $\sigma_{wp}=0.4$ and $q_s=0.8$. The results are presented for four values of the activity strength $f_A$, including the passive mixture, $f_A=0$. 
(b) Same as (a), but on a semi-log plot. These are from DFT calculations.}
\label{fig_s9}
\end{figure}

\subsection{DFT Results: Phase Diagram} 
\label{section2}

The DFT calculations outlined in Sections~\ref{subsection1.3}-\ref{subsection1.5} are most conveniently performed in the grand-canonical ensemble~\cite{egorov2021}. A useful control parameter is provided by the polymer reservoir packing fraction $\eta_{p}^{r}=(\pi\sigma_{p}^{3}/6)\exp(\mu_p/(k_BT))$. In order to construct the binary mixture phase diagram under spherical confinement, we define the dimensionless grand potential density $\beta\omega(r)$ as the integrand of the right-hand side of Eq.~(\ref{omega}). The phase diagram is then obtained numerically by imposing the following two requirements: 1) the colloid chemical potential $\mu_c$ has the same value (within the prescribed tolerance of $10^{-6}$) in the two coexisting phases, and 2) the reduced grand potential $\beta \Omega^{*}=3\int_{0}^{R} dr r^2 \beta\omega(r)/R^3$ has the same value in the two coexisting phases within the same tolerance.

In order to impose these coexistence conditions, we perform two sequences of DFT calculations starting with two different values of the colloid chemical potential -- one above and one below the corresponding bulk coexistence value of the passive mixture, which is known from prior DFT work~\cite{egorov2021}. The colloid chemical potential is then incremented in small positive/negative steps, respectively, and at each step of these two sequences the equilibrium density profiles in the spherical cavity are calculated and the reduced grand potential is obtained. The resulting values of $\beta\Omega^{*}$ in the two phases are plotted as functions of $\beta\mu_{c}$, and the point where the two lines cross corresponds to the phase coexistence in the spherical cavity. These calculations are performed for a range of the polymer reservoir packing fractions ($0.65\le\eta_{p}^{r}\le 1.1$) and for several values of the activity strength $f_A$, including the passive mixture, $f_A=0$. The resulting phase diagrams in the variables ($\eta_{p}^{r}$, $\beta\mu_c$) are shown in Fig.~\ref{fig_s8} for four values of the activity strength $f_A$. One sees that with increasing the activity strength, the coexistence curve moves to lower values of the colloid chemical potential (for any given value of $\eta_{p}^{r}$). 

In order to facilitate a comparison with the MD simulation results, the DFT phase diagram is replotted in the variables of polymer and colloid packing fractions in Fig.~\ref{fig_s9}.  One sees that with increasing the activity strength $f_A$, the two-phase region widens considerably, in agreement with the MD results.

\end{document}